\documentclass{ws-ijmpb}

\begin{document}

\markboth{A. Siddiki and R.~R. Gerhardts}
{Disorder Effects in the Screening Theory of the IQHE}


%

\catchline{}{}{}{}{}

%


\title{RANGE-DEPENDENT DISORDER EFFECTS ON THE\\
 PLATEAU-WIDTHS CALCULATED  WITHIN \\
THE SCREENING THEORY OF THE IQHE  }

\author{AFIF SIDDIKI}

\address{Physics Department, Arnold Sommerfeld Center
of Theoretical Physics,  and CeNS,\\
Ludwigs-Maximilians-Universit{\"a}t M{\"u}nchen,\\
Theresienstr. 37, D-80333 M\"unchen, Germany\\
siddiki@theorie.physik.uni-muenchen.de }

\author{ROLF R GERHARDTS}

\address{Max-Planck-Institut f\"ur Festk\"orperforschung,\\
  Heisenbergstr. 1 D-70569 Stuttgart, Germany\\
R.Gerhardts@fkf.mpg.de}

\maketitle

\begin{history}
\received{25 August 2006}



\end{history}

\begin{abstract}
We summarize the screening theory of the integer quantized Hall effect (IQHE)
and emphasize its two key mechanisms: first, the existence, in certain
magnetic field intervals, of incompressible strips, with integer values of the
local filling factor and quantized values of longitudinal and Hall
resistivity, and second, the confinement of an imposed dissipative current to
these strips, leading to the quantization of the global resistances. We
demonstrate that, without any localization assumption, this theory
explains the enormous experimental reproducibility of the quantized resistance
values, as well as experimental results on the potential distribution in narrow
Hall bars. We further demonstrate that inclusion of long-range potential
fluctuations allows to apply the theory to wider Hall bars, and can lead to a
broadening of the quantum Hall plateaus, whereas short-range disorder tends to
narrow the plateaus.
\end{abstract}


\section{Introduction}

Twenty-five years after its discovery, one should believe that  the integer
quantized Hall effect (IQHE),\cite{vKlitzing80:494}
observed on two-dimensional electron systems (2DES) in high magnetic fields,
is well understood. Indeed it is often ``explained'' as a consequence of
Landau quantization and localization of electronic states, within a single
particle picture that considers the
Coulomb interaction between the electrons as irrelevant.
This ``explanation''  ignores, however, important aspects
of the IQHE, e.g.\ the enormous accuracy with which the quantized resistance
values can be reproduced experimentally.\cite{Bachmair03:14} It also can not
describe, or  even  contradicts,  recent experiments by E. Ahlswede {\em et
  al.}\cite{Ahlswede01:562} on
the Hall potential distribution  in narrow Hall bars (width
$\lesssim 20 \,\mu$m).
A major purpose of the present paper is to emphasize that both, these recent
experiments and the enormous experimental reproducibility of the quantized
resistance values, can be understood within a conventional local transport
theory, which takes into account the peculiar screening, i.e.\ Coulomb
interaction, effects in 2DES under high magnetic fields, but avoids any
localization assumptions. Since some details of this theory have already been
published,\cite{Siddiki04:195335,Siddiki04:3541} we focus here on the key
mechanisms of this approach. Moreover, we add a discussion of long-range
potential  fluctuations, which become important in wider Hall bars and
indicate a possible extension of our approach. The systems we
have in mind contain a 2DES in a GaAs/(AlGa)As heterostructure, populated by
ionized (Si-)donors, which are   distributed randomly in a plane parallel to
that of the 2DES.

\section{Importance of screening induced edge profile for the IQHE}
\subsection{Bulk resistivity and accuracy of the IQHE}
If we admit that a bulk theory for the resistivity of a homogeneous system,
like that of localization,\cite{Kramer03:172} yields quantization
in a regime of filling factors, say around $\nu=2$, the edge regions of a
finite sample, where the filling factor drops to zero, must lead to errors.
If we estimate the error by the ratio of (width of edge region) over (sample
width), a typical depletion length of about 100 nm and an accuracy of
$10^{-8}$ or better requires a sample width of ten meters or more.
Thus, a theory for homogeneous samples cannot explain  the accuracy of
  the IQHE in samples of realistic size.
\subsection{Edge states and screening}
The sample edges are taken into account in the B\"uttiker picture, which
considers an upwards bending of the Landau energy levels as a consequence of
the confining potential.\cite{Buettiker88:9375}
The relevant, current-carrying  edge states are believed
to  be the states of the Landau bands immediately at the Fermi edge. This
picture traces the IQHE back to the conductance quantization in quasi-1D
systems. However, in the quasi-1D situation all channels (Landau bands)
 carry the same amount
of current, which seems not to be true in the Ahlswede
experiment.\cite{Ahlswede01:562}

This edge-state picture has been criticized theoretically by Chklovskii {\em
  et al.}\cite{Chklovskii92:4026} who
 argue that, for a smooth confinement potential, it leads to a
step-like density profile, with plateaus corresponding to integer filling
factors. To change the smooth density profile at vanishing magnetic field,
  $B=0$,   into   this $B$-dependent step-profile would cost a lot of Coulomb
  energy.
{\em Screening} will avoid such large and energetically expensive changes of
 the density profile.
{\em Compressible strips} with nearly perfect screening will occur, in which a
partially filled Landau level (LL) is pinned to the Fermi
energy, so that at zero temperature the total potential within the strip is
flat, while the density varies very similar to the $B=0$ case.
Adjacent compressible strips will be separated by {\em incompressible strips}
  (ISs).
There the Fermi level is between neighboring LLs, the density is constant, and
  the potential varies by the amount of a cyclotron energy
across the IS. We assume here and in the following spin
degeneracy.

Some questions remain, which we have to answer:
Do all these ISs really exist? Where does the current flow ?
\subsection{The Ahlswede experiment on the Hall potential distribution}
Important answers were given by the experiments of Ahls\-wede and
cowor\-kers,\cite{Ahlswede01:562,Ahlswede02:165,Weitz00:247,Weitz00:349}
who used a  scanning force microscope to  measure the local
electrostatic force across a narrow Hall bar. From this they calculated the
Hall potential, which determines the local current distribution.
They observed the quantized Hall effect (near filling factors $\nu=2$ and
$\nu=4$) and, related to this,
three different types of  potential drop across the Hall
bar:\cite{Ahlswede01:562}
{\em (I)}, a linear drop for $B$-fields well outside a QH plateau;
{\em (II)}, a non-linear drop in the center region, for $B$-fields near the
upper edge of a QH plateau.
{\em (III)}, for $B$-fields well within a plateau  a potential profile was
observed which is constant in the center and drops across strips, which move
with decreasing $B$ towards the edges of the Hall bar.
In all cases {\em only one current carrying strip} at each edge is observed,
at the position expected for the innermost {\em incompressible strip}$\,$!

Earlier calculations,\cite{Pfannkuche92:7032,Oh97:13519}
 which describe an imposed {\em dissipationless} Hall
current along a Hall bar, can not explain these different types of potential
drop. Apparently we need a {\em local transport theory} that is able to
describe
{\em dissipative currents} in the presence of
 compressible and {\em incompressible strips}.

\section{Screening theory of the IQHE}

\subsection{Local distribution of dissipative currents in the IQH regime}

We consider a simplified Hall bar in the stripe $|x|\leq d$ of the
$x$-$y$-plane, and assume translation invariance in $y$-direction.
We impose a fixed total current $I_0=\int_{-d}^d dx\, j_y(x,y)$ along the Hall
bar and write for the
current density ${\bf j}$ Ohm's law with a local resistivity tensor
$\hat{\rho}$ and
a driving electric field ${\bf E}$, which we assume to be the gradient of the
position-dependent electrochemical potential $\mu^{\star}$,
\begin{equation}
 \hat{\rho} ({\bf   r})\, {\bf j}({\bf r})={\bf  E}({\bf
 r})\equiv  {\bf \nabla} \mu^{\star} ({\bf   r}) /e \,,
\qquad \hat{\rho}({\bf r})=\big[\hat{\sigma}\big( n_{\rm el}
  ({\bf r})\big)\big]^{-1}. \label{eq1}
\end{equation}
The resistivity tensor, with components
$\rho_{yy}=\rho_{xx}=\rho_{\rm l}$ and  $\rho_{xy}=-\rho_{yx}=\rho_{\rm
 H}$, and its inverse, the conductivity tensor $\hat{\sigma}$, are assumed to
depend only on the local electron density $ n_{\rm el}(x)$.
 The dissipative current density vanishes in
thermodynamic equilibrium, since then $\mu^{\star}$ is constant over the
sample.

In the stationary, translation-invariant case, one has
$ {\bf \nabla} \cdot  {\bf j}({\bf r}) = 0$ and
${\bf \nabla} \times  {\bf  E}({\bf   r})= {\bf 0}$, and, with $\partial_y j_y
=0$ and $\partial_y E_x=0$, one sees $ \partial_x j_x =0$ and
$\partial_x E_y=0$, so that $j_x(x) \equiv 0$ and $ E_y(x)=E_y^0$
independent of $x$. With this and Eq.~(\ref{eq1}) one immediately obtains
the components of current density and driving electric field,
\begin{equation}
 E_y(x)\equiv E_y^0\,, \quad j_y(x)=\frac{1}{\rho_{\rm
 l}(x)}\,E_y^0\,, \quad E_x(x)=\frac{\rho_{\rm H}(x)}{\rho_{\rm
l}(x)}\,E_y^0 \,, \label{eq2}
\end{equation}
in terms of the resistivity components.
Knowing the field components, it is easy to calculate the position dependence
of the electrochemical potential,
\begin{equation}
\mu^*(x,y)=\mu^*_0 +e E_y^0 \,\left\{ y+
  \int_0^x dx' \,
      \frac{\rho_{\rm  H}(x')}{\rho_{\rm l}(x')} \right\}\,.
\end{equation}

If a model for the local conductivity tensor
$\hat{\sigma}( n_{\rm el}(x))$ is given, it is straightforward to
calculate current distribution and electrochemical potential.
For instance in the Drude model there exists no longitudinal
magnetoresistance, i.e.\
$\rho_{\rm l}(x)=m/[e^2 \tau n_{\rm el}(x)]$ is independent of $B$, and $
\rho_{\rm H}(x)=\omega_c \tau \rho_{\rm l}(x)$. Thus, $j_y(x)$ is proportional
to the electron density $ n_{\rm el}(x)$ and the Hall field $ E_x(x)$ is
independent of $x$, i.e.\ the potential varies linearly across the sample
({\em type I} behavior).

\subsection{Current confinement to incompressible strips}

As a consequence of Landau quantization, a (spin-degenerate) 2DES at (even)
integer filling factor $\nu=n$ is, in the limit of vanishing temperature ($T
\rightarrow 0$), inert in the sense that no elastic scattering is possible
(since occupied and unoccupied electron states are separated by the cyclotron
energy $\hbar \omega_c \gg k_BT$). In the absence of scattering, the
resistivity components take on the free-electron values, $\rho_{\rm l}(\nu
\rightarrow n) \rightarrow 0$, $\rho_{\rm H}(\nu \rightarrow n) \rightarrow
h/(n \cdot e^2)$. The essence of our local approach is the assumption that
this remains valid on (sufficiently wide) incompressible strips with integer
{\em   local} filling factor,
\begin{equation} \nu(x)=n\,:
 \quad \rho_{\rm l}(x)   \rightarrow 0\,, \quad
\rho_{\rm H}(x) \rightarrow h/(n \cdot e^2) \quad \mbox{for}\quad T \rightarrow
0\,.  \label{essence}
\end{equation}
If such  ISs exist, the integral  $\int_{-d}^d  dx\,[1 / \rho_{\rm
    l}(x)] \rightarrow \infty$ diverges for $T   \rightarrow 0$. Since the
    total imposed current $I_0=\int_{-d}^d dx\,[E_y^0 / \rho_{\rm
    l}(x)] $ is fixed, this implies
    $E_y^0 \rightarrow 0$, and the current density tends to zero outside the
    incompressible strips, where $ {\rho_{\rm l}(x)}$ remains non-zero.
From Eq.~(\ref{eq2}) we see also that the Hall field $E_x(x)$ vanishes outside
    the ISs, so that the Hall potential drops only across these strips, i.e.\
    shows the {\em type III} behavior observed in the Ahlswede experiment.

If only incompressible strips with the same integer value $n$ of the local
filling factor exist, the Hall resistivity takes on the same value $h/(n \cdot
e^2)$ wherever the Hall field remains finite, and can be taken out of the
integral, $V_{\rm H}=\int_{-d}^d dx\,E_x(x) \rightarrow  [h/(n \cdot e^2)]
 I_0$. As a result, the {global resistances are quantized}:
\begin{equation}
R_{\rm l} \propto E_y^0 \rightarrow 0, \quad\quad R_{\rm H}
  \rightarrow h/(n \cdot e^2)\,. \label{quantization}
\end{equation}
The resistances are quantized, since all the current flows in the ISs. Then
dissipation and entropy production vanish, i.e.\ assume their minimum possible
values, in accordance with the rules of irreversible thermodynamics.
The deviation from the quantized values becomes {\em exponentially small} with
decreasing temperature [$\sim \exp(-\hbar \omega_c /2 k_BT)$], as a consequence
of Landau quantization and Fermi statistics, without any localization
assumptions.
\subsection{Application to the Ahlswede experiment}
To apply this formalism to the experimental situation, we have to choose a
model for the conductivity tensor as function of the filling factor. We have
considered two models which yield very similar results, first, an approach
based on a Gaussian approximation of
the collision-broadened Landau levels,\cite{Guven03:115327,Gerhardts75:285}
and, second, the self-consistent Born
approximation (SCBA) to the electron-impurity
problem.\cite{Siddiki04:195335,Ando74:959,Ando82:437} The latter allows
 for a fully consistent treatment of Landau
level broadening and  magneto-conductivity tensor.

To obtain a position-dependent conductivity tensor, we replaced in these
models the filling factor by the local filling factor of the simplified Hall
bar, calculated under due consideration of screening effects.  For given
density profile $n_{\rm el}(x)$ of the 2DES, we had to solve
Poisson's equation under reasonable electrostatic boundary conditions. We
adopted a model used by Chklovskii {\em et
  al.},\cite{Chklovskii92:4026,Chklovskii93:12605} which assumes all charges
and gates in a single plane and takes advantage of the methods of complex
analysis. However, whereas these authors  calculate $n_{\rm el}(x)$ in the
``electrostatic approximation'' from the crude assumption of perfect screening
in the 2DES, we used a self-consistent Thomas-Fermi-Poisson
approximation\cite{Lier94:7757,Oh97:13519}  (TFPA), which describes the
properties of the 2DES more realistically by its density of states (DOS).
To calculate $n_{\rm el}(x)$ in Thomas-Fermi approximation, can be
justified from a more reliable Hartree-type calculation, if the
potential $V(x)$ confining the electrons to the Hall bar is so smooth that it
is
nearly constant over the extent of  occupied energy eigenfunctions. If this
condition is not satisfied, the TFPA may yield incompressible strips, which
do not exist if the finite extent of the wavefunctions is taken into account.
We found, in agreement with earlier work by Suzuki and
Ando,\cite{Suzuki96:46,Suzuki98:415}
that Hartree-type calculations are considerably more restrictive in predicting
ISs than the TFPA.
For more details see Fig.1 of Refs.\refcite{Siddiki04:195335} and
\refcite{Siddiki04:3541}, and the related discussions.

Our Hartree-type calculations show that, for the sample widths and electron
densities of interest, incompressible strips exist only in magnetic field
intervals of finite width (the plateau regions of the IQHE), and that in these
intervals  only a single IS exists on each side of the sample, with the same
integer value of the filling factor on both sides. This is in agreement with
the Ahlswede experiment.

For a direct comparison with the experiment, we should realize that
the scanning force microscope measures the (gradient of the ) electro{\em
  static}  potential $V(x)$, not the electro{\em chemical} potential
$\mu^{\star}(x)$ , which we can
easily calculate for given $\hat{\sigma} (n_{\rm el}(x))$. To calculate the
feedback of the imposed dissipative current on the density profile and the
electrostatic potential, we assume {\em local equilibrium} in the stationary
dissipative state, i.e.\ we repeat the equilibrium calculation for the
position-dependent electrochemical potential and iterate the procedure with
the new density profile until convergence for density profile, electrostatic
and electrochemical potential is obtained. For a high imposed current $I_0$,
this yields a considerable change of $n_{\rm el}(x)$.\cite{Guven03:115327}
For a weak  $I_0$, however, the density change is small and the
current-induced change of $V(x)$  has practically the same
position dependence as $\mu^{\star}(x)$. Therefore, in the linear response
regime it is sufficient to calculate $\mu^{\star}(x)$ from the equilibrium
density profile and to compare its position dependence with the measured
change of $V(x)$.

\subsection{Simulation of non-local corrections}
Our  local transport model in connection with the TFPA leads to two types of
problems. We have already discussed the effect of the finite extent of
wavefunctions, which prohibits ISs in edge regions, where density profile and
confinement potential are too steep.
As a second failure, the local relation between the statistically defined
quantities like
current density, resistivity tensor and gradient of the electrochemical
potential will break down on scales comparable with the average distance
between electrons. It also leads to artifacts like a singular current density
at positions where the local filling factor assumes integer values, even if
there no IS of finite width exists.
We can get rid if both problems in a very simple way. We calculate the density
profile $n_{\rm el}(x)$ within the TFPA, with this the conductivity tensor
$\hat{\sigma}\big(\nu(x)\big)\equiv \hat{\sigma}(x)$, and then simulate
non-local effects by  coarse-graining the conductivity
  tensor according to
\begin{equation}
   \hat{\bar{\sigma}}(x)=\frac{1}{2\lambda}\int_{-\lambda}^{\lambda}
    d\xi \, \hat{\sigma}(x+\xi),   \qquad \mbox{with} \quad
     \lambda \sim \lambda_F/2, \label{coarse}
\end{equation}
where $\lambda_F$ is the Fermi wavelength.
With the coarse-grained $\hat{\bar{\sigma}}(x)$, we calculate
  current distribution and $\mu^{\star}(x)$.

\subsection{Summary of  results}
Having emphasized the basic ingredients and the key mechanisms of our approach
to the understanding of the IQHE in narrow Hall bars, we now summarize the
basic results.\cite{Siddiki04:195335,Siddiki04:3541}
For narrow GaAs Hall bars (width $\lesssim 15\,\mu$m) containing a
(spin-degenerate) 2DES of
typical density [$n_{\rm el}(0)\lesssim 4\cdot 10^{11}\,$cm$^{-2}$] we find
non-overlapping magnetic field intervals of finite width, in which
incompressible strips with a well defined (even) integer value of the local
filling factor exist. For $B$ values in these intervals, the imposed
dissipative current is, with decreasing
temperature, increasingly confined to these incompressible strips, so that
the global longitudinal and Hall resistances approach the local
resistivity values on these ISs. The deviation of the Hall resistance from the
quantized values decreases exponentially with decreasing temperature.
The intervals under discussion are thus the plateau regimes of the IQHE.

For the simple, translation invariant Hall bars, which we have considered so
far, the
electron density (at $B=0$) decreases monotonously from the center towards the
edges. As $B$ decreases from high values, ISs occur first in the center,
broaden rapidly, and then
split into two strips at a lower $B$ value. With further decreasing
$B$, the strips move  towards the edges, shrink, and finally disappear at a
$B$ value considerably higher than the one at which the IS with the next
higher integer filling factor occurs in the center.

The widths of the ISs, and as a consequence the widths of the quantum Hall
plateaus, shrink with increasing
temperature,\cite{Lier94:7757} with increasing value of the
phenomenological coarse-graining parameter $\lambda$ ($\sim
20$-$50\,$nm), which we use to simulate non-local effects, finite width of
wavefunctions, etc., and with increasing collision broadening of the Landau
levels.\cite{Siddiki04:195335}
For reasonable parameter values, the calculated position-dependence of the
Hall potential shows all the different types of profiles
observed in the Ahlswede experiment.

\section{Long-range potential fluctuations}
 \begin{figure}[tb]
 \centerline{\psfig{file=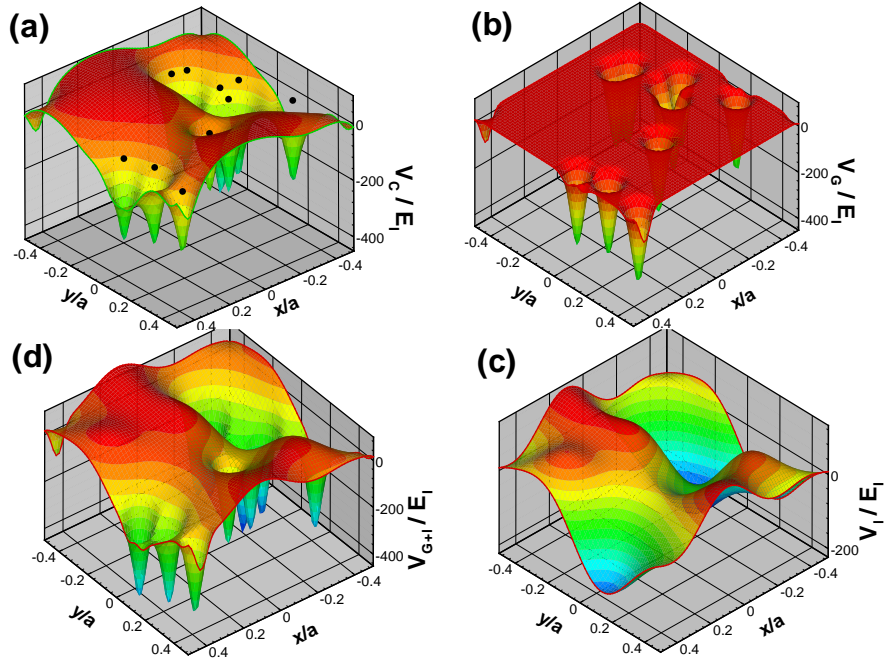,width=4.8in}}
 \caption{Potential of $N_I=10$ randomly distributed donors (black
 dots) in the unit cell of a
 square lattice with lattice constant $a=3\,\mu$m, at distance $z=90\,$nm
 from the  2DES, $\bar{\kappa}=12.4$. (a) Coulomb potentials; (b)  Gaussian
 approximation; (c)
 long-range part given by Fourier coefficients with ${\bf q}=2\pi(n_x,n_y)/a$
 for $|n_x|\leq 2$ and   $|n_y|\leq 2$; (d) sum of (b) and
 (c). Average potentials are  substracted in (a) - (d), energy unit
 $E_I=\pi \hbar^2\cdot N_I/(ma^2)$.
\label{donorpot-qha}
 }
 \end{figure}
\subsection{Separation of short- and long-range potential fluctuations}

So far we have considered impurity scattering, within the SCBA,
in the calculation of conductivities and level broadening of the Landau DOS.
In principle we should consider randomly distributed Coulomb scatterers at a
distance $z$ from the 2DEG. However, since the SCBA does not describe coherent
multi-center scattering, the overlapping long-range parts of the Coulomb
potentials would lead to unphysically large damping effects. As usual we have,
therefore, taken into account only the short-range parts of the potentials,
approximated by Gaussians, i.e.\ we have replaced
\begin{equation}
V_{\rm imp}({\bf r};z)\!=\!\sum_{j} \frac{-e^2/\bar{\kappa}}{\sqrt{({\bf
      r\!-\!r}_j)^2+z^2}} \quad \mbox{by} \quad
V_{\rm Gauss}
({\bf r};z)\!=\!\sum_{j} \frac{-e^2}{\bar{\kappa} |z|}\, \exp \big(\!-\!
      \frac{({\bf r\!-\!r}_j)^2}{2 z^2}\big), \label{gauss-appr}
\end{equation}
where ${\bf r}$ and ${\bf r}_j$ are vectors within the $x$-$y$-plane.
To get a feeling for the short- and long-range parts of the potential
fluctuations,
we have plotted in the upper left panel of Fig.~\ref{donorpot-qha}
the potential created by ten donors
distributed randomly in a square.
To the right, we show the superposition of the corresponding Gaussians, i.e.,
the short-range part (SRP).
Below we plot the long-range part (LRP), defined by the lowest order Fourier
coefficients.
In the lower left panel we show the sum of long-range and short-range parts,
which looks qualitatively very similar to the original superposition of
Coulomb potentials.

The SRP and the LRP of the fluctuating donor potential do
not only look very different, they also behave very differently concerning both
their dependence on the distance $z$ between donor layer and 2DES and on the
screening by the 2DES. The SRP is dominated by large Fourier coefficients and
therefore decreases rapidly (exponentially) with increasing $z$, however is
only weakly screened. The LRP, on the other hand, is dominated by low-order
Fourier coefficients, thus decreases only slowly with increasing $z$, but is
strongly screened by the 2DES. We take this as justification to treat SRP and
LRP very differently. For the SRP we neglect screening effects, use the
Gaussian approximation and consider it within the SCBA for the calculation of
conductivities and level broadening, as we did before.
To take the LRP into account, we add it to the external confinement potential
and treat it together with this in the self-consistent screening calculations.

 \begin{figure}[h]
 \centerline{\psfig{file=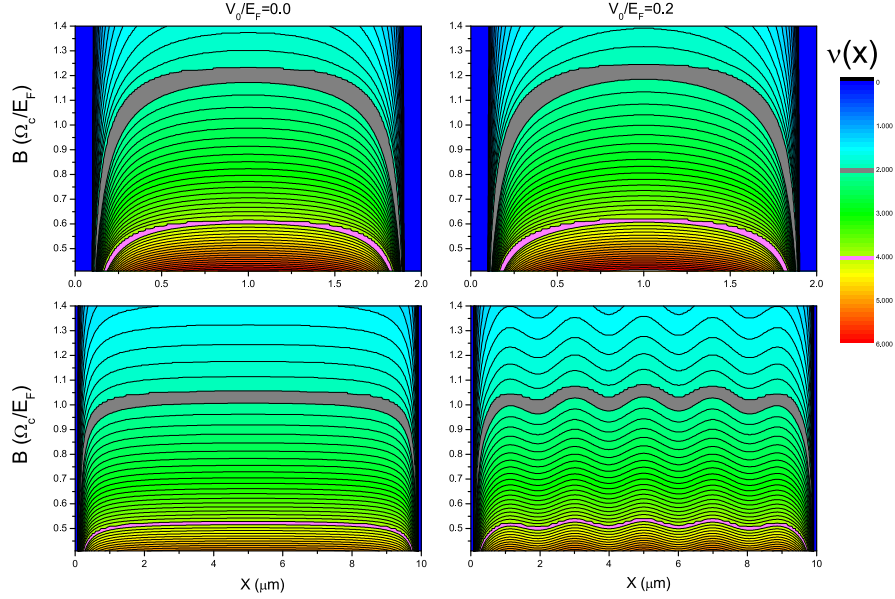,width=5.0in}}
 \vspace*{-8pt}
 \caption{Color-coded plot of the local filling factor versus position $x$ and
 cyclotron energy $\Omega_c=\hbar \omega_c$  without (left panels) and with
 (right panels) external
 modulation potential $V_{\rm mod}(x)=-V_0 \sin (p\, \pi x/2d)$ in $0\leq x
 \leq 2d$, for $2d=2\, \mu$m and $p=1$ (upper) and $2d=10\, \mu$m and $p=9$
 (lower panels); gray indicates $\nu(x)=2$ and pink $\nu(x)=4$. Average
 densities: $2.88\cdot 10^{11}\,$cm$^{-2}$ (upper) and $3.69\cdot
 10^{11}\,$cm$^{-2}$ (lower panel); $k_BT/E_F=0.02$, $V_0/E_F=0.2$; at $T=0$,
 $B=0$, in all cases
 the width of the symmetric density profile is $2d-200\,$nm.
\label{lrp-model}
 }
 \end{figure}

\subsection{Effect on the plateau width of the IQHE}
In Fig.~\ref{lrp-model} we simulate long-range disorder by a simple harmonic
modulation potential of the same period in a narrow and a five times wider
Hall bar. Without the modulation, the density profile is flatter (i.e.\
screening is more effective) in the wider sample, so that the Fermi energy
$E_F=\bar{n}_{\rm el}/D_0$, which yields the average filling factor
$\bar{\nu}=2$ at $\hbar \omega_c /E_F =1$, is only slightly smaller than the
energy $E_F^0= n_{\rm el}(x=d, B=0,T=0)/D_0$, which determines the
high-magnetic-field edge of the $\nu=2$ quantum Hall plateau (QHP) at $\hbar
\omega_c /E_F^0 =1$.

\begin{figure}[h]
\begin{minipage}[b]{0.48\linewidth} \centering
\includegraphics[width=0.97\linewidth]{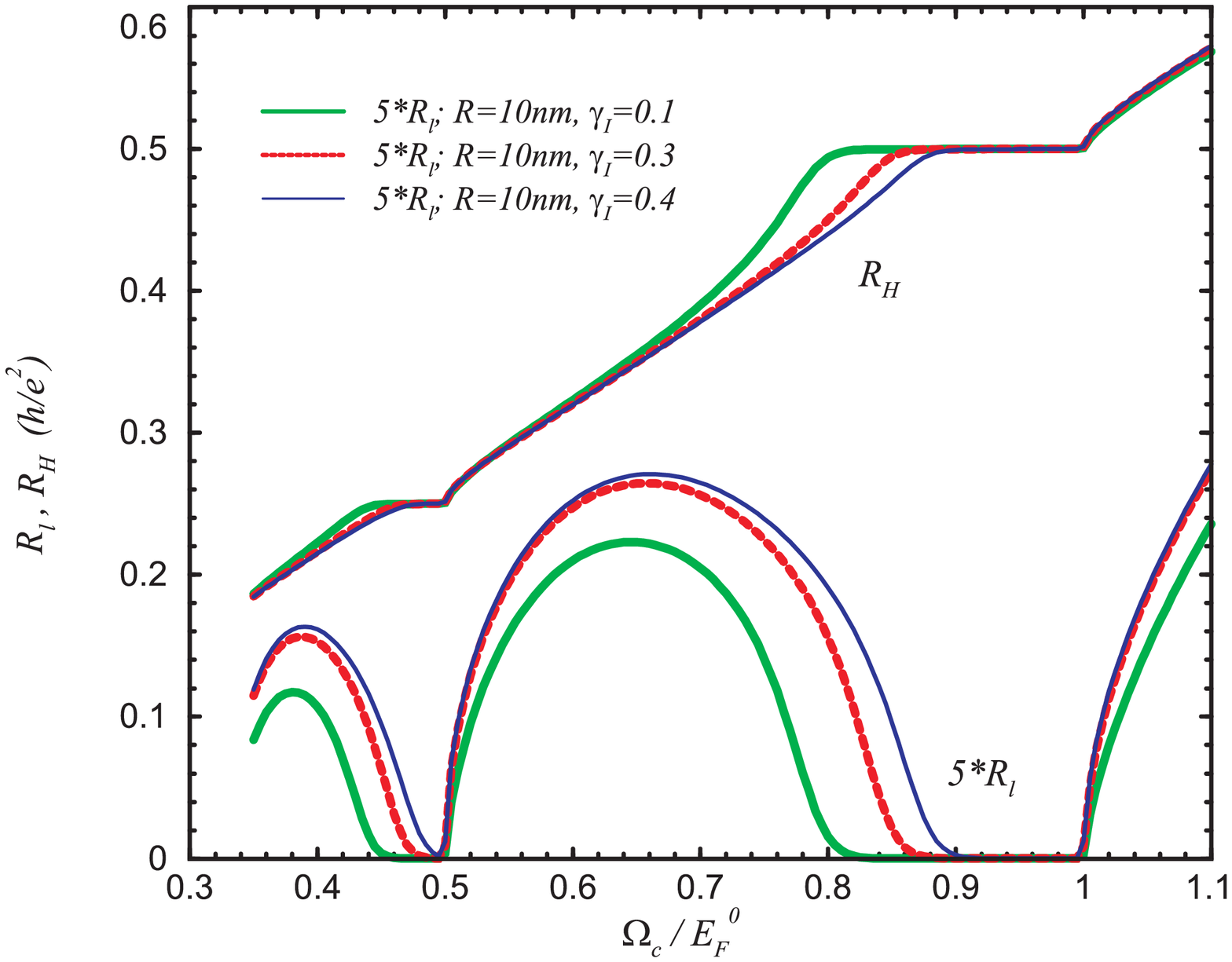}\\
\caption{Dependence of the resistance curves on the strength $\gamma_I$ of
  short-range disorder. For details see
  Ref.~\protect\refcite{Siddiki04:195335}.  \label{srf} }
\end{minipage}
\hfill
\begin{minipage}[b]{0.48\linewidth} \centering
\includegraphics[width=0.95\linewidth]{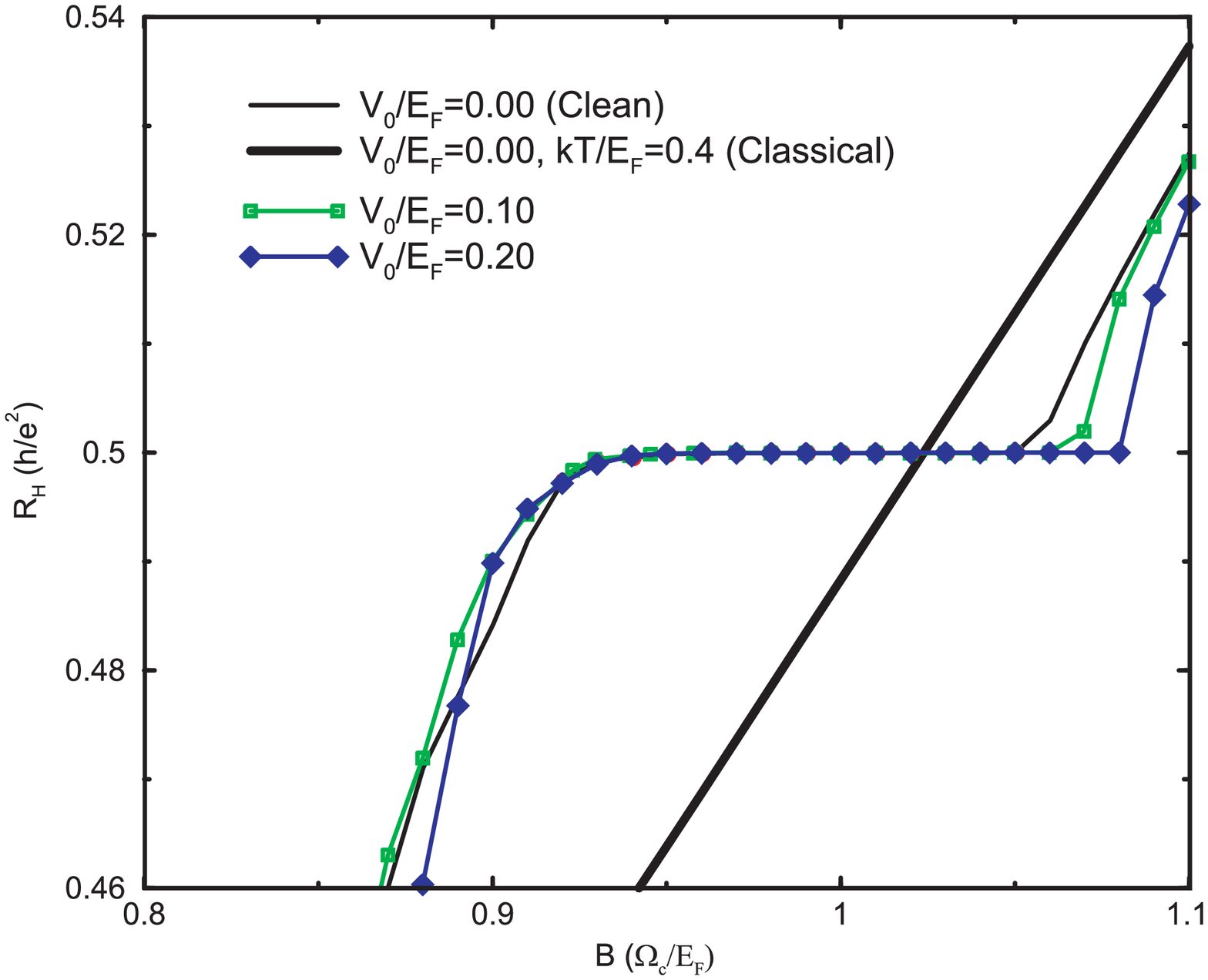}\\
\caption{Dependence of the $\nu$=2 Hall plateau on the amplitude $V_0$ of
  long-range disorder for the sample of width $2d=10\,\mu$m of
  Fig.~\protect\ref{lrp-model}.  \label{lrf} }
\end{minipage}
\end{figure}

For the narrow sample, the modulation potential has little effect. It
slightly enhances the confinement potential, thus increases the electron
density in the center ($x=d$), and shifts the high-$B$ edge of the QHP to
slightly higher $B$-values. For the wider sample and sufficiently large
modulation amplitude, at certain $B$-values additional incompressible strips
may occur inside the sample. Thus, increasing amplitude of long-range disorder
may lead to a broadening and stabilizing of quantum Hall plateaus, as opposed
to an increase of short-range disorder, which leads to an increase of the
Landau level width, i.e.\ a decrease of the Landau gaps and, thereby, to a
shrinking of the plateaus. These different effects of short- and long-range
disorder are demonstrated in Fig.~\ref{srf} and Fig.~\ref{lrf}, respectively.
Figure~\ref{srf} is calculated for Gaussian potentials, like in
Eq.~(\ref{gauss-appr}), with a range $R=\sqrt{2}\,z$, and $\gamma_I$ is
proportional to the density and the square of the potential amplitude of the
impurities.

\section{Conclusion}

Long-range potential fluctuations across wide Hall bars may
produce additional incompressible strips, and thereby stabilize the quantum
Hall plateaus, shift them to higher $B$ fields, and broaden them.
Thus, the screening theory of the IQHE, which explains
  the experimental   results on narrow Hall bars very well, has also potential
  for application to   wide samples.
Although this theory in its present state contains a number of
phenomenological assumptions, which in the future may be justified or
abandoned, we believe that it contains the clue for the understanding of the
IQHE and its astonishing reproducibility even in narrow samples: In the
plateau regime of the IQHE the current is forced to flow through channels in
the sample, where the quantization conditions hold locally. Screening is a
 mechanism to provide, in inhomogeneous systems and in magnetic field
intervals of finite width, such channels in the form of incompressible stripes
or, more generally, percolating incompressible regions. Localization effects
may play a role in macroscopic
samples, but they seem not to be relevant in the narrow samples considered
here.


\section*{Acknowledgements}

We are grateful to E. Ahlswede, J. Weis, F. Dahlem, J. G. S. Lok, and K. von
Klitzing for stimulating discussions.

\section*{References}


\end{document}